\documentclass[12pt]{article}

\newcommand{\sect}[1]{\setcounter{equation}{0}\section{#1}}

\def\be{\begin{equation}}
\def\ee{\end{equation}}
\def\ba{\begin{eqnarray}}
\def\ea{\end{eqnarray}}

\title{{\bf Classical and Thermodynamic Stability of Black Branes}}

\author{{\bf Harvey S. Reall} \\ Physics Department, 
Queen Mary College \\ Mile End Road, London E1 4NS, United Kingdom}

\date{8 April 2001 \\ Preprint QMUL-PH-01-06}

\begin{document}

\maketitle

\begin{abstract}

It is argued that many non-extremal black branes exhibit a classical
Gregory-Laflamme instability if, and only if, they are locally thermodynamically
unstable. For some black branes, the Gregory-Laflamme
instability must therefore disappear near extremality. 
For the black $p$-branes of the type II supergravity theories, 
the Gregory-Laflamme instability disappears near extremality for
$p=1,2,4$ but persists all the way down to extremality for $p=5,6$ 
(the black D3-brane is not covered by the analysis of this paper).
This implies that the instability also vanishes for the near-extremal 
black M2 and M5-brane solutions.

\end{abstract}

\sect{Introduction}

\subsection{The Gregory-Laflamme instability}

The Gregory-Laflamme (GL) instability is a classical instability of
non-extremal black branes \cite{gregory:93, gregory:94}. The existence
of this instability is often explained by arguing that a black $p$-brane 
has a lower entropy per unit $p$-volume than an array of black
holes with the same total mass and charge per unit volume\footnote{
Assuming that the black holes can carry the charge of the black
brane. If not, then one might instead consider an array of uncharged
black holes threaded by an extremal black brane.}, 
and must therefore be unstable. 
However, this explanation leaves something to be desired because it
uses thermodynamic (and therefore quantum) arguments to explain a
classical phenomenon.

It is, of course, well-known that the classical laws of black hole
mechanics \cite{bardeen:73} can be re-interpreted as thermodynamic
laws once it is appreciated that black holes emit thermal radiation
\cite{hawking:75} and therefore have a temperature and associated
entropy. Conversely, the laws of thermodynamics applied to black 
holes can sometimes be reformulated in classical
language. The entropy argument above would become an argument
based on the second law of black hole mechanics \cite{hawking:73},
which states that the area of the event horizon is non-decreasing in
any physical process. Since the area of the horizon would increase if
a black brane were to evolve to an array of black holes, the second
law certainly permits a classical instability of black
branes. However, it does not {\it require} it.

\subsection{The Gubser-Mitra conjecture}

The GL instability has recently been re-examined by Gubser and Mitra (GM)
\cite{gubser:00a, gubser:00b} from the perspective of the AdS/CFT correspondence
\cite{maldacena:98}. It was argued that the absence of tachyonic modes
in the AdS glueball calculations \cite{witten:98, csaki:99} is
evidence that near-extremal D3-branes are in fact classically stable,
and that this is consistent with the work of GL because GL only
checked that an instability exists for branes that are far from
extremality. However, this appears to contradicts the claim of GL
\cite{gregory:95} that their instability persists all the way down 
to extremality\footnote{
That is, for any non-zero value of the non-extremality parameter. It was
argued that the instability is absent for extremal branes, which is
as one would expect in a supersymmetric theory.}. 
Furthermore, to apply AdS/CFT arguments
one must first take the decoupling limit \cite{maldacena:98} of the
black brane, which makes it only infinitesimally
non-extremal\footnote{Thanks to Simon Ross for emphasizing this point to
me. Throughout this paper, ``near-extremal'' will mean that 
some non-extremality parameter lies in a {\it finite}
neighbourhood of its extremal value.}.

More interestingly, the following conjecture was made by GM: {\it a
black brane with a non-compact translational symmetry is classically
stable if, and only if, it is locally thermodynamically stable.} 
The definition of local thermodynamic stability adopted by GM was the
one used in \cite{gubser:99}, namely that the Hessian of the entropy
with respect to the other extensive thermodynamic variables 
(i.e. mass and charges) should be negative definite. This is the
mathematical statement of the condition that it shouldn't be possible
for the system to gain entropy by splitting into several parts whilst
conserving energy and charge, i.e., it is a statement about the
microcanonical ensemble.

At first sight, the GM conjecture appears to contradict the claim
\cite{gregory:95} that the GL instability persists all the way down
to extremality. For example, near-extremal D3-branes have positive
specific heat, which implies that they satisfy the above definition of
local thermodynamic stability and should therefore be classically
stable according to the conjecture. However, the work of GL only
covered black branes carrying {\it magnetic} charge with respect to a
{\it Neveu-Schwarz} $n$-form. 
Therefore, the only black brane of type II supergravity
covered by their work is the NS5-brane. This brane does indeed have
negative specific heat all the way down to extremality, so the fact
that it is classically unstable agrees with the GM conjecture. One
might think that the work of GL could be extended to the F-string by
dualizing the NS 3-form, but this changes the coupling to the dilaton,
taking one outside the class of theories considered by GL. One of the
results of the present paper will be to show that the GL instability occurs
for many more solutions than just the ones considered by GL.

The condition that there should be some non-compact translational
symmetry was included in the GM conjecture to exclude black holes 
from the scope of the conjecture. This is because it is well-known 
that many black holes, for example Schwarzschild, 
are classically stable \cite{vishveshwara:70}
yet have negative specific heat and are therefore locally
thermodynamically unstable.

The GM conjecture can be regarded as a refinement of the entropy argument
mentioned in the first paragraph of this paper. This entropy argument
refers to {\it global} thermodynamic stability. It seems much more
plausible that, if there is a connection between classical and
thermodynamic stability, it should be local thermodynamic
stability that is relevant. For example, one might imagine some kind
of potential barrier separating the black brane from an array of black
holes, so the only way that the brane could gain in entropy would be
by quantum tunneling through this barrier. In that case, since the
instability would be quantum mechanical in nature, global
thermodynamic instability would not imply classical instability. 

The evidence that GM gave for their conjecture was based on studying
black holes solutions \cite{duff:99} of gauged ${\cal N}=8$ supergravity in
four dimensions. These solutions are asymptotic to
anti-de Sitter space and the black holes 
considered were large compared with the AdS radius of
curvature. This presumably implies that the curvature of their
horizons should not be significant so they would be expected to exhibit
similar behaviour to black branes. Therefore if the conjecture is
correct, then it should apply to these black holes too. 
It was found that a classical instability
appeared almost precisely where local thermodynamic stability was
lost, with a small discrepancy arising from numerical error. Note that
large asymptotically flat black holes would not be expected to behave
like black branes because there is no length scale with which the
curvature of the horizon can be compared. 

\subsection{Proving the conjecture}

In this paper, a semi-classical proof of the GM conjecture will be
given, using the Euclidean path integral approach to quantum
gravity. This has been succesful in providing a
semi-classical explanation of other connections between classical and
thermodynamic properties of black holes, for example the relationship
between the entropy of a black hole and the area of its horizon
\cite{gibbons:77}.
It is convenient to review some basic facts about this approach
\cite{hawking:79}. The Euclidean path integral for the canonical
ensemble is
\be
 Z = \int d[{\bf g}] \exp ( - I[{\bf g}] ),
\ee
where $I$ denotes the Euclidean Einstein-Hilbert action. The path
integral is taken over Riemannian manifolds $(M,{\bf g})$ that are
asymptotically flat, i.e., tend to the flat metric on $S^1 \times R^3$
outside some compact region. The $S^1$ should have proper length
$\beta = 1/T$, where $T$ is the temperature. 

This path-integral is only well-defined in the semi-classical
approximation. In this approximation, one looks for saddle-points of
the Euclidean action, i.e., solutions of the Euclidean Einstein
equations. These are referred to as gravitational instantons. Near a
gravitational instanton, the metric can be written as
\be
 g_{\mu\nu} = \bar{g}_{\mu\nu} + \delta g_{\mu\nu},
\ee
where $\bar{g}_{\mu\nu}$ is the metric of the instanton, and
$\delta g_{\mu\nu}$ denotes the perturbation. One can add gauge fixing
and ghost terms to the action to deal with the gauge freedom in
$\delta g_{\mu\nu}$. The action can be expanded around the saddle point:
\be
 I[{\bf g}] = I_0 [\bar {\bf g}] + I_2 [\bar{\bf g}, \delta {\bf g}],
\ee
where $I_2$ is quadratic in the fluctuation. The contribution to $I_2$
from the trace of the metric perturbation is a wrong-sign kinetic
term.  This is the well-known conformal factor problem
\cite{gibbons:78}. Nothing new will be said about this problem in the
present paper. The part of $I_2$ involving the traceless part of the
metric perturbation is, in transverse gauge, proportional to
\be
 \int d^4 x \sqrt{\bar{g}}  H^{\mu\nu} \Delta_L H_{\mu\nu},
\ee
where $H_{\mu\nu}$ is transverse traceless and $\Delta_L$ is the Euclidean
Lichnerowicz operator. The path-integral over the transverse traceless part of 
the metric perturbation can be evaluated by expanding $H_{\mu\nu}$ in
eigenfunctions of $\Delta_L$, i.e., solutions of
\be
\label{eqn:lich}
 \Delta_L H_{\mu\nu} = \lambda H_{\mu\nu}.
\ee
The boundary conditions are that $H_{\mu\nu}$ should vanish fast
enough at infinity to have finite norm, and be regular everywhere else.
The contribution to the partition function from the instanton is
\be
 A \exp (-I_0[\bar{\bf g}]),
\ee
where $A$ is a functional determinant that arises from the
path-integrals over the metric perturbation and ghosts. In particular,
it includes a factor of $(\det \Delta_L)^{-1/2}$. This implies that if
$\Delta_L$ has a negative eigenvalue then there will be an imaginary
contribution to the path integral. This pathology arises from the fact
that the canonical ensemble for gravity is not well-defined in
infinite volume \cite{hawking:76, hawking:79}, i.e., it is an
indication of a thermodynamic instability. 

For the Euclidean Schwarzschild solution, there is a single transverse traceless
negative mode \cite{page:78, gross:82}, i.e., a solution to equation \ref{eqn:lich}
with negative $\lambda$. The eigenvalue is \cite{gross:82}
\be
 \lambda = -0.19 M^{-2},
\ee
where $M$ denotes the Schwarzschild mass. The existence of this
negative mode implies that the Euclidean Schwarzschild solution is
merely a saddle point of the Euclidean action, rather than a local
minimum \cite{page:78}. Therefore this instanton cannot be regarded as giving an
approximation to the partition function. The correct interpretation is
as an instanton describing a non-perturbative instability of flat
space at finite temperature \cite{gross:82}. 

It is now possible to explain the main idea of this paper. 
Consider an uncharged black $p$-brane in $4+p$ dimensions with metric
\be
 ds^2 = g_{\mu\nu} dx^{\mu} dx^{\nu} + \delta_{ij} dz^i dz^j,
\ee
where $g_{\mu\nu}$ denotes the four dimensional Lorentzian Schwarzschild metric
and $z^i$ denote the flat spatial worldvolume directions of the brane.  
GL considered \cite{gregory:93} metric perturbations of this solution of the
form
\be
 h_{\mu\nu} = \exp{(i \mu_i z^i)} H_{\mu\nu}, \qquad h_{\mu i} =
 h_{ij} = 0,
\ee
where transverse traceless gauge is used for $h_{\mu\nu}$,  and $H_{\mu\nu}$ is
spherically symmetric. GL looked for modes that are regular on the
future horizon of the brane and decay at infinity, but grow
exponentially in time: $H_{\mu\nu} \propto \exp(\Omega t)$.
The equation of motion for the perturbation is \cite{gregory:93}
\be
\label{eqn:lich2}
 \tilde{\Delta}_L H_{\mu\nu} = -\mu^2 H_{\mu\nu},
\ee
where $\tilde{\Delta}_L$ denotes the four dimensional Lorentzian 
Lichnerowicz operator and $\mu^2 = \sum_i \mu_i \mu_i$.

GL showed that there is a critical value $\mu_*$ such that for every $\mu <
\mu_*$, equation \ref{eqn:lich2} has a solution that is regular on the
future horizon and decays at spatial infinity, but grows exponentially in
time. For $\mu > \mu_*$, no growing mode was found and the solutions
are oscillatory in time. It follows that the solution with $\mu =
\mu_*$ must be independent of time. This solution will be referred to as
the {\it threshold unstable mode.} Note that the classical second law
implies that a black brane compactified on a sufficiently small torus
cannot evolve to an array of black holes, which suggests that the
black brane should be stable against small wavelength (large $\mu$)
perturbations, and therefore that a time-independent mode must exist
if the brane is unstable to large wavelength perturbations. So one can argue
that the existence of such a mode is necessary for a classical
instability without referring to the numerical results
of \cite{gregory:93, gregory:94}. This will be important for the more
general black brane solutions considered later in this paper.

It is now possible to make contact with the Euclidean path integral by Wick
rotating $H_{\mu\nu}$ for the threshold unstable mode. Since
$H_{\mu\nu}$ is static and spherically symmetric, Wick rotation
simply changes the sign of $H_{tt}$.
$H_{\mu\nu}$ is translationally invariant on
the brane world-volume so it can be dimensionally reduced to
four dimensions. Wick rotation transforms $\tilde{\Delta}_L$ into
$\Delta_L$, so equation  \ref{eqn:lich2} turns into equation
\ref{eqn:lich} with $\lambda = -\mu_*^2$.  Hence it appears that the
classical Lorentzian threshold unstable mode corresponds precisely to
a Euclidean negative mode. It will be shown later that the Wick rotated
mode does indeed satisfy the correct boundary conditions for a
negative mode. 

This argument can be checked by comparing the numerical results of
\cite{gross:82} and \cite{gregory:93}. The latter gave results for
horizon radius $r_+ = 2$ \cite{gregory:94}, which corresponds to
$M=1$, i.e., $\lambda = -0.19$. This would give $\mu_* = 0.44$, which
is in excellent agreement with figure 1 of \cite{gregory:93}. It is
possible to extend this matching to the other dimensions for which the
negative mode has been calculated \cite{prestidge:00c}, namely
$\lambda = -1.60 r_+^{-2}$ in five dimensions and  $\lambda = -3.39
r_+^{-2}$ in seven dimensions. For $r_+=2$, this gives $\mu_* = 0.63$ and $\mu_* =
0.92$ respectively, again in good agreement with \cite{gregory:93}. A
simple scaling argument shows that this agreement must extend to
all values of $r_+$. 

It is clear that the classical GL instability of the black brane
is very closely related to the thermodynamic instability of the
Schwarzschild black hole. The threshold unstable mode is related to
the Schwarzschild negative mode by discarding the $\exp( i \mu_i z^i)$
factor and Wick rotating\footnote{
This argument shows how a negative mode of the Schwarzschild black
hole is related to a static zero mode of the black string. This
relationship was noted in \cite{gregory:88}, although the connection
with classical stability was not appreciated (indeed it was argued
that the black string is classically stable!)}. 
The existence of negative modes is closely
linked to local thermodynamic stability, as will be explained shortly. It
therefore seems likely that the above argument can be generalized into
a proof of the GM conjecture for more general black brane
solutions. That will be the subject of the rest of this paper.

\subsection{Outline of this paper}

The next section describes the class of black branes that will be
discussed in this paper, namely magnetically charged dilatonic black
branes. In section \ref{sec:negative}, the
relationship between local thermodynamic stability and negative modes
is explained. Section \ref{sec:classical} discusses classical
perturbations and in section \ref{sec:quantum} it is shown how these 
are related to negative modes. The final section summarizes the main
conclusions and suggests directions for future research. 

\sect{Black brane solutions}

\label{sec:solns}

This paper will discuss black $p$-brane solutions of the action
\be
 I = \epsilon \int d^D x \sqrt{\epsilon g} \left(-R + \frac{1}{2}
(\partial \phi)^2 + \frac{1}{2n!} e^{a\phi} F_{(n)}^2 \right),
\ee
where $\epsilon = +1$ for Euclidean signature and $-1$ for Lorentzian
signature. $F_{(n)}$ denotes a $n$-form field strength obtained from a
$(n-1)$-form potential $A_{(n-1)}$ by $F_{(n)} = dA_{(n-1)}$.

Black brane solutions of this action have been given by several
authors \cite{horowitz:91, guven:92, duff:94, duff:96}. The notation
of \cite{duff:96} will be used here. The magnetically charged black brane
solutions can be written:
\ba
 ds^2 &=& \left(1+\frac{k}{r^{\tilde{d}}} \sinh^2 \mu \right)^{-\frac{4
 \tilde{d}}{\Delta (D-2)}} \left( \epsilon e^{2f} dt^2 + dz^i dz^i \right)
 \nonumber \\ &+&
 \left(1+\frac{k}{r^{\tilde{d}}} \sinh^2 \mu \right)^{-\frac{4
 d}{\Delta (D-2)}} \left( e^{-2f} dr^2 + r^2 d\Omega_{\tilde{d}+1}^2
 \right), \nonumber \\
 e^{-\frac{\Delta}{2a} \phi} &=& 1+\frac{k}{r^{\tilde{d}}} \sinh^2 \mu ,
 \qquad e^{2f} = 1-\frac{k}{r^{\tilde{d}}}.
\ea
The parameters $d$, $\tilde{d}$ and $\Delta$ are defined by
\be
\label{eqn:paramdef}
 \tilde{d} = n-1, \qquad d=p+1= D-\tilde{d}-2, \qquad \Delta = a^2 +
 \frac{2d\tilde{d}}{D-2}.
\ee
The $n$-form field strength is a constant times the volume form on
$S^n$. The two free parameters $k$ and $\mu$ are related to the charge
$\lambda$ and mass per unit $p$-volume $m$ by \cite{duff:96}
\be
 \lambda = \frac{\tilde{d}k}{\sqrt{\Delta}} \sinh 2\mu, \qquad m = k
 \left(\frac{4 \tilde{d}}{\Delta} \sinh^2 \mu + \tilde{d}+1 \right).
\ee
The temperature is
\be
 T = \frac{\tilde{d}}{4\pi r_+} \left(\cosh \mu\right)^{-4/\Delta}.
\ee
The extremal limit is $k \rightarrow 0$, $\mu \rightarrow \infty$
with $k e^{2\mu}$ fixed. It is straightforward to verify that the
specific heat at fixed charge is always negative if $2
\tilde{d}/\Delta \le 1$. However, if $2 \tilde{d}/\Delta > 1$ then this
specific heat is negative far from extremality but positive near
extremality.

For the black $p$-branes of the type II supergravity theories
\cite{horowitz:91}, with the exception of the D3-brane, 
one can always dualize to make the solution magnetically charged. The
solutions then take the above form, and one finds that the specific
heat changes sign near extremality when $p \le 4$ (i.e., D0, F1, D1,
D2, D4) but not for $p=5,6$ (NS5, D5, D6) (D$p$-branes with $p>6$ will
not be considered here because they are not asymptotically flat). 
For the solutions considered by
Gregory and Laflamme, the specific heat never changes sign. The
specific heat of the D3-brane does change sign but this brane lies
outside the class of solutions considered in this paper. Nevertheless,
it seems likely that the arguments of the present paper could be generalized
to include this case. 
For the black M2 and M5 brane solutions \cite{guven:92} of eleven
dimensional supergravity, the specific heat is negative far from
extremality but becomes positive near extremality.

Dimensional reduction on the worldvolume of the above solution leads
to a magnetically charged dilatonic black hole solution in $n+2$
dimensions. The properties of such solutions were extensively
discussed in \cite{gibbons:88}. It is clear that the temperature of
this black hole must be the same as the temperature of the black
brane. It is also the case that the charge of the black hole is the
same as the charge of the black brane, and the mass of the black hole
is proportional to the mass per unit $p$-volume of the black brane
\cite{lu:93}. It follows that the specific heat at constant charge of
the black hole has the same sign as the specific heat at constant
charge of the black brane. 

\sect{Negative modes and the specific heat}

\label{sec:negative}

\subsection{Canonical ensemble}

The canonical ensemble for gravity can be rendered well-defined by
imposing boundary conditions at a finite boundary rather than at
infinity \cite{york:86}. Consider a spherical cavity of surface area
$4\pi r_B^2$, whose boundary is held at a fixed temperature $T$. 
The partition function in the canonical ensemble is given by a path integral over
Riemannian geometries $(M,{\bf g})$ such that $\partial M = S^1 \times
S^2$, where the $S^1$ has proper length $\beta = 1/T$ and the $S^2$ has
area $4 \pi r_B^2$.

In the case of a pure gravity theory, for small $T r_B$ the only
solution to the Euclidean Einstein equations that obeys these boundary
conditions is flat space with a periodically identified time
coordinate. For larger $T r_B$, there are three solutions: flat space,
a small Schwarzschild black hole, and a large Schwarzschild black hole
(with size of order $r_B$). The smaller black hole has negative
specific heat and the larger black hole has positive specific heat. 

It is straightforward to demonstrate that the black hole with negative
specific heat cannot be a local minimum of the Euclidean action and
must therefore have a negative mode. For a Schwarzschild black hole,
this was shown in \cite{prestidge:00} using a 
construction given in \cite{whiting:88}. The idea is
to construct a 1 parameter family of geometries for which the
Euclidean action (relative to flat space) takes the form
\be
\label{eqn:actiondecomp}
 I = \beta E(x) - S(x),
\ee
where $x$ is the parameter labelling the path. 
There should be should be some value $x = x(T)$ 
for which the geometry becomes the black hole geometry of
interest. At this value, $E$ and $S$ are the energy and entropy of the
black hole:
\be
 E(x(T)) = E(T), \qquad S(x(T)) = S(T).
\ee
For other values of $x$, the interpretation of the functions $E$
and $S$ is not important. The black hole extremizes the action, so at 
$x = x(T)$, 
\be
 \left( \frac{\partial I}{\partial x} \right)_T = 0,
\ee
which is simply the condition for thermal equilibrium of the black
hole with a heat bath of temperature $T$:
\be
 \left[\beta \frac{dE}{dx} - \frac{dS}{dx}\right]_{x
= x(T)} = 0 \Rightarrow \left[\frac{dE}{dS} \right]_{x = x(T)} = T.
\ee
The second derivative of the action along the path parametrized by
$x$ is
\be
 \left( \frac{\partial^2 I}{\partial x^2} \right)_T =
 \left( T \frac{dx}{dT} \right)^{-2} \frac{dE}{dT}.
\ee
It is evident from this equation that if the specific heat $dE/dT$ is 
negative then the black hole cannot be a local minimum of the action
and therefore it must have a negative mode\footnote{
Note that the path of off-shell geometries parametrized by $x$ does
not correspond directly to an eigenfunction of the Lichnerowicz
operator, but rather to some linear combination of
eigenfunctions. Since the action decreases along this path, at least
one of these eigenfunctions must have a negative eigenvalue.}. 

In order to exhibit this connection between specific heat and negative
modes it was necessary to assume that there exists a 1-parameter family
of geometries, including the black hole, for which equation
\ref{eqn:actiondecomp} is obeyed. This will now be demonstrated for the
black holes of relevance to this paper, namely those obtained by
dimensional reduction of the black brane solutions given above. These are
dilatonic black holes in $n+2$
dimensions carrying magnetic charge with respect to a $n$-form field
strength. The construction is a straightforward generalization of
\cite{whiting:88}. The Euclidean metric can be written
\be
\label{eqn:metric}
 ds^2 = U(r) d\tau^2 + V(r)^{-1} dr^2 + R(r)^2 d\Omega_n^2.
\ee
The exact form of $U(r)$, $V(r)$ and $R(r)$ will not be important. The
dilaton is a function of $r$ alone, and the $n$-form is proportional
to the volume form on $S^n$. 

In the canonical ensemble, if the boundary is at $r=r_B$ then the
proper length of the Euclidean time direction at $r=r_B$ should be $\beta =
1/T$. It is convenient to normalize $\tau$ so that $\tau \sim \tau
+ \beta$, which implies $U(r_B)=1$. The functions $U(r)$ and $V(r)$
both vanish at the horizon, $r=r_+$. The condition for regularity at $r=r_+$ is
\be
\label{eqn:reg}
 \sqrt{U'(r_+) V'(r_+)} = \frac{4\pi}{\beta},
\ee
which gives $r_+ = r_+(T)$. The black hole solution is specified by
the parameters $k$ and $\mu$, which are determined by the temperature
$T$ and charge $\lambda$. This black hole will be referred to as $B(T,\lambda)$.

The other geometries in the family will be
taken to have a metric of the form \ref{eqn:metric} but now with
$U(r)$ treated as an arbitrary function. $V(r)$, $R(r)$, the dilaton and
the $n$-form are all taken to be the same as for a black hole with
charge $\lambda$ but arbitrary temperature. In other words, $k$ and
$\mu$ are constrained only to keep $\lambda$ fixed but not $T$, so this gives
a 1-parameter family of metrics on the surfaces of constant
$\tau$. This family can be parametrized by the value of
$r_+$. Choosing $r_+ = r_+(T)$ reproduces the geometry of the spatial
sections of $B(T,\lambda)$. 

This choice of geometry for surfaces of constant $\tau$ 
implies that the constraint equations of
general relativity are satisfied on these surfaces. 
$\tau$ is still taken to be identified with
period $\beta$. The only restrictions on $U(r)$ are $U(r_B)=1$ (so the 
metric obeys the boundary conditions), $U(r_+)=0$ (so the topology is
the same as that of the black hole $B(T, \lambda)$), that $U'(r_+)$ obeys equation
\ref{eqn:reg} (so that the metric is regular at $r=r_+$) and that
$U(r)$ is positive for $r>r_+$. It is obviously possible to choose 
a suitable family of functions $U(r)$ so that the metric becomes that
of $B(T, \lambda)$ when $r_+ = r_+(T)$.

The action of these geometries (relative to flat space) can be calculated 
using the methods of \cite{hawking:79, hawking:96}.
One first excises the region $r < r_+ +
\epsilon$. In the region $r>r_+ + \epsilon$, one write the action in
Hamiltonian form. The contribution from this region is then simply
\be
 I_+ = \beta H,
\ee
where $H$ is the Hamiltonian evaluated on the above metric. Since the
above metric is static and satisfies the constraint equations, the
only contribution to $H$ comes from a surface term at $r=r_B$, giving
\be
 I_+ = \beta E,
\ee
where $E$ depends only on $r_+$ and the charge $\lambda$. 
The total action is given by adding
the action of the region $r< r_+ + \epsilon$. In the limit $\epsilon
\rightarrow 0$, this is given entirely by a gravitational surface
term\footnote{
A surface term involving the $n$-form vanishes for magnetic black holes.}
at $r=r_+$ which, after using equation \ref{eqn:reg} reduces to
\be
 I_- = -\frac{A}{4G},
\ee
where $A$ is the area of the horizon at $r=r_+$, and $G$ is Newton's constant.
The total action is therefore
\be
 I = I_+ + I_- = \beta E - S,
\ee
where $S = A/4G$, which depends only on $r_+$. This establishes 
equation \ref{eqn:actiondecomp} (with $x = r_+$).
The proof that negative specific heat implies the existence of a 
negative mode then proceeds as above. The only subtlety is that, since
the charge was held fixed above, the specific heat should be
calculated at constant charge. The reason
for working in the canonical ensemble (fixed charge) rather than the
grand canonical ensemble (fixed potential)
is that the theory does not contain charged particles and
therefore the charge cannot fluctuate. This is also true for
electrically charged black holes\footnote{
In the case of electric charge (for $n=2$), it is necessary to add an extra
surface term to the Euclidean action when working in the canonical
ensemble \cite{mann:95, hawking:95}.}. 

\subsection{Physical interpretation}

There is a simple physical interpretation of the off-shell geometries
considered above. Consider a typical member of this family.
The geometry of the spatial sections is simply
that of a black hole of charge $\lambda$ but with temperature $T' \ne
T$. Since the metric is static (independent of $\tau$)
analytic continuation to Lorentzian signature at a surface of
constant $\tau$ yields initial data for solving the Lorentzian
Einstein equations and the unique solution is the black hole with
charge $\lambda$ and temperature $T'$. In other words, the path of
off-shell geometries constructed above 
corresponds to an instability in which the black hole changes
temperature at fixed charge, which is precisely the instability that
one would expect for a black hole in contact with a heat
bath. Note that the off-shell geometry is {\it not} simply given by analytic
continuation of the metric of a black hole of temperature $T'$ since
this would not give a metric with the correct period for the Euclidean
time at $r=r_B$. Note also, that if this interpretation is correct
then many different $U(r)$ correspond to the same Lorentzian black
hole. 

It has been shown that negative specific heat implies the existence of
a negative mode. To see that this is a physical negative mode (rather
than a negative mode arising from the conformal factor problem),
consider the off-shell geometries near $B(T,\lambda)$. These can be
regarded as perturbations of $B(T,\lambda)$. The function
$U(r)$ is essentially arbitrary for $r_+<r<r_B$, so by adjusting
$U(r)$, the trace of the metric perturbation can be freely
adjusted. If the negative mode were associated with the
conformal factor problem then one would expect increasing the
trace of the metric perturbation to make the action more negative. 
However, since the constraint equations are satisfied on
surfaces of constant $\tau$, the action does not depend on $U(r)$ and
therefore this change in the trace of the metric perturbation does not
change the action. It follows that the negative mode whose existence
was demonstrated above is a non-conformal negative mode. 

The above discussion considered black holes in finite cavities. This
is a rather unphysical situation since it is hard to see how an
appropriate cavity could be constructed. The rest of this paper will
consider only black holes in infinite space, i.e., the $r_B
\rightarrow \infty$ limit will be assumed. 

\subsection{Other ensembles}

Since the charge of the black hole cannot fluctuate, the condition for local
thermodynamic stability in the microcanonical ensemble is that the
second derivative of the entropy with respect to the mass must be
negative. Equivalently, the specific heat should be positive. So the
condition for local thermodynamic stability of the black hole is the
same in the canonical and microcanonical ensembles.

In a theory containing charged
particles, one might choose to use the grand ensemble for electrically
charged holes. This could lead to new negative modes involving the
gauge field. For example, consider the Reissner-Nordstrom
solution. The necessary and sufficient conditions for local
thermodynamic stability in the grand ensemble is that the Hessian of 
the Gibbs free energy with respect to the entropy $S$ and charge $Q$ be positive
definite. This is equivalent to\footnote{
See section 21 of \cite{landau:80} and replace $-V$ with $Q$ and $p$
with $\Phi$.}
\be
 \left( \frac{\partial E}{\partial T} \right)_Q > 0, \qquad \left(
 \frac{\partial \Phi}{\partial Q} \right)_T > 0,
\ee
where $\Phi$ denotes the potential at infinity (or at $r=r_B$) in a
gauge regular at the horizon. In four dimensions, one finds that the
former equation is violated for $Q/M < \sqrt{3}/2$ but the latter
equation is violated for $Q/M > \sqrt{3}/2$. Thus one would expect an
instability analagous to the thermodynamic instability of the
Schwarzschild black hole for small $Q$, and the associated negative
mode would involve the metric as above. However, for large $Q$, the
instability involves the black hole gaining or losing charge, so the
instability would involve the electromagnetic field and charged matter
fields, as for the black holes studied in \cite{gubser:00a,
gubser:00b}. Hence, in this case one would expect a negative mode involving the
electromagnetic and matter fields. In a theory that did not contain
charged matter, one would work in the canonical ensemble and the
second of the above stability conditions would not be required so
one would expect no negative mode for $Q/M > \sqrt{3}/2$. This is
confirmed by numerical calculations \cite{prestidge:00b}, which show
that the negative mode involving the metric vanishes precisely at $Q/M
= \sqrt{3}/2$.  

Note that it has only been shown that negative specific heat implies the
existence of a negative mode. Conversely, the existence of a negative
mode implies that the black hole is only a saddle point of the
Euclidean action, which implies that it suffers from some pathology in
the canonical ensemble. This suggests that it should be locally
thermodynamically unstable. The Reissner-Nordstrom example is evidence 
that this is true, but I am unaware of a more general proof. 

\sect{Classical stability}

\label{sec:classical}

The aim of this section is to derive the equations of motion for
classical perturbations of the black brane solutions discussed in
section \ref{sec:solns}. 

\subsection{Equations of motion}

The analysis will be simplified by a convenient choice of conformal
frame. For the moment the choice of conformal frame will be kept
arbitrary by writing
\be
 \hat{g}_{MN} = e^{2b\phi} g_{MN},
\ee
where a caret refers to the $D$-dimensional Einstein
metric. In the new frame, the action is
\be
 I = \epsilon \int d^D x \sqrt{\epsilon g} \left[ e^{-\beta\phi} \left(-R + k
(\partial \phi)^2 \right)  + \frac{1}{2n!} e^{\alpha \phi} F_{(n)}^2
\right],
\ee
where
\be
\label{eqn:betadef}
 \beta = -(D-2) b, \qquad k = \frac{1}{2} - (D-1)(D-2)b^2, \qquad
 \alpha = a + (D-2n)b.
\ee
The equations of motion can be written
\be
 \nabla_M \left(e^{\alpha \phi} F^{M N_1 \ldots N_{n-1}} \right) = 0,
\ee
\be
 \nabla^2 \phi - \beta (\partial \phi)^2 = \frac{a}{2n!} e^{(\alpha +
\beta) \phi} F^2,
\ee
\ba
 R_{MN} &=& (k+\beta^2) \partial_M \partial_N \phi - \beta \nabla_M
 \nabla_N \phi \nonumber \\
 &+& \frac{1}{2(n-1)!} e^{(\alpha + \beta) \phi} F_{M P_1
\ldots P_{n-1}} F_N{}^{P_1 \ldots P_{n-1}} - \frac{\Lambda}{4n!}
e^{(\alpha + \beta)\phi} F^2 g_{MN},
\ea
where
\be
 \Lambda = \left(\frac{D-2n}{D-2}\right) \left( \frac{2\beta^2}{D-2} -1
\right) + \frac{2\alpha \beta}{D-2} + 1. 
\ee
The conformal frame that will be used is given by 
\be
\label{eqn:framechoice}
 b = \frac{\tilde{d}}{a (D-2)}.
\ee
In this frame, there is no conformal factor multiplying the flat
worldvolume directions of the black brane metric. Making this choice has
the added advantage that it gives
\be
 \Lambda = 0.
\ee
Perturbing the equations of motion gives
\ba
\label{eqn:deltaFeq}
 \nabla_N \left(e^{\alpha \bar{\phi}} \delta F^{N P_1 \ldots P_{n-1}}
 \right) + (\alpha \partial_N \delta \phi) e^{\alpha \bar{\phi}} 
 \bar{F}^{N P_1 \ldots P_{n-1}} -(n-1) e^{\alpha \bar{\phi}}
\bar{F}^{NQ [P_2 \ldots P_{n-1}} \nabla_N h^{P_1]}_Q \nonumber \\
 - h^N_M \nabla_N
\left( e^{\alpha \bar{\phi}} \bar{F}^{M P_1 \ldots P_{n-1}} \right) -
\nabla_N \left(h^N_M - \frac{1}{2} h \delta^N_M \right) e^{\alpha \bar{\phi}} 
 \bar{F}^{M P_1 \ldots P_{n-1}} = 0.
\ea
\ba
 \nabla^2 \delta \phi &-& h^{MN} \nabla_M \nabla_N \bar{\phi} + \beta
 h^{MN} \partial_M \bar{\phi} \partial_N \bar{\phi} - \partial_M
 \bar{\phi} \nabla_N \left( h^{MN} - \frac{1}{2} h \bar{g}^{MN} \right) -
 2\beta \partial \bar{\phi} \cdot \partial \delta \phi \nonumber \\
&+& \frac{a}{2(n-1)!} e^{(\alpha + \beta) \bar{\phi}} h^{MN} \bar{F}_{M P_1
 \ldots P_{n-1}} \bar{F}_N{}^{P_1 \ldots P_{n-1}} - \frac{a}{n!}
 e^{(\alpha + \beta) \bar{\phi}} \bar{F} \cdot \delta F \nonumber \\
 &-& \frac{a}{2n!}
 e^{(\alpha+\beta)\bar{\phi}} (\alpha+\beta) \delta \phi \bar{F}^2 = 0.
\ea
\ba
 \nabla^2 h_{MN} &-& \nabla_M \nabla^P \left( h_{PN} - \frac{1}{2} h
 \bar{g}_{PN} \right) - \nabla_N \nabla^P \left( h_{PM} - \frac{1}{2} h
 \bar{g}_{PM} \right) \nonumber \\
 &-& 2R_{P(M}h^M_{N)} - 2R_{(M|PQ|N)} h^{PQ} + 2\beta
\left(\nabla_{(M} h^P_{N)} - \frac{1}{2} \nabla^P h_{MN} \right)
\partial_P \bar{\phi} \nonumber \\
&-& 2\beta \nabla_M \nabla_N \delta \phi + 4 (k+\beta^2) \partial_{(M}
\bar{\phi} \partial_{N)} \delta \phi \nonumber \\
 &+& \frac{1}{(n-1)!} e^{(\alpha + \beta)\bar{\phi}} \left[2 \delta
F_{(M}{}^{P_1 \ldots P_{n-1}} \bar{F}_{N) P_1 \ldots P_{n-1}} - (n-1)
h^{PQ} \bar{F}_{MP R_1 \ldots R_{n-2}} \bar{F}_{NQ}{}^{R_1 \ldots
R_{n-2}} \right. \nonumber \\ &+& \left. 
 (\alpha + \beta) \delta \phi \bar{F}_{M P_1 \ldots P_{n-1}}
\bar{F}_{N}{}^{P_1 \ldots P_{n-1}} \right] = 0,
\ea
where bars denote background quantities, and the background metric is
used to define covariant derivatives and raise and lower indices. 

It is convenient to split the $D$-dimensional indices $M,N$ into indices
$i,j$ along the spatial worldvolume directions and indices $\mu,\nu$
in the other directions. Indices $m,n$ will be used to
denote directions on the $n$-sphere. The choice of conformal frame
made above implies that connection components and curvature components
vanish if they have an $i$ index. $z$ will be used to denote spatial
worldvolume coordinates and $x$ to denote other coordinates. 

\subsection{The Gregory-Laflamme ansatz}

Following Gregory and Laflamme \cite{gregory:94}, only s-wave
perturbations will be considered. For such perturbations,
\be
 h^{tm} = h^{rm} = h^{im} = 0, \qquad h^m_n = K(t,r)\delta^m_n.
\ee
Other components, and the dilaton perturbation, are also taken to be
independent of the coordinates on the sphere. This implies that the
terms involving $h_{MN}$ and $\delta \phi$ in equation
\ref{eqn:deltaFeq} vanish, just as in \cite{gregory:94}. 
It is therefore consistent to set $\delta F = 0$. 
In \cite{gregory:94}, it was shown that there was no unstable mode
associated with $\delta F$. It seems very likely that the same is true
here, so $\delta F = 0$ will be assumed from now on. 

GL made the ansatz
\be
 h_{MN} = \exp(i \mu_i z^i) H_{MN}(x), \qquad \delta \phi = \exp(i \mu_i z^i) f(x)
\ee
and considered only longitudinal perturbations, i.e.,
\be
 H_{i \mu} = i \mu_i A_{\mu} (x),
\ee
\be
 H_{ij} = \mu_i \mu_j H(x).
\ee
The same ansatz will be adopted here. 
Under gauge transformations, the metric perturbation transforms as
\be
 h_{MN} \rightarrow h'_{MN} = h_{MN} - \nabla_M \xi_N - \nabla_N
 \xi_M,
\ee
so under a gauge transformation with $\xi_M = \exp(i \mu_i z^i) \Xi_M$
and $\Xi_i = i \mu_i \Xi$ the metric perturbation becomes
\ba
 H'_{\mu\nu} &=& H_{\mu\nu} - \nabla_{\mu} \Xi_{\nu} - \nabla_{\nu}
 \Xi_{\mu}, \nonumber \\ 
 A'_{\mu} &=& A_{\mu} - \partial_\mu \Xi - \Xi_\mu, \\
 H' &=& H + 2\Xi. \nonumber
\ea
The quantity $H_{\mu\nu} - 2 \nabla_{(\mu}A_{\nu)} -
\nabla_{\mu}\nabla_{\nu} H$ is gauge invariant under such
transformations. Note that it is possible to transform away $A_{\mu}$
and $H$ by a gauge transformation. The following gauge choice can
therefore be made
\be
\label{eqn:pertgauge}
 H_{i \mu} = H_{ij} = 0.
\ee
This is {\it not} the same gauge as used by Gregory and
Laflamme. However, in many respects, it is much
simpler to use this gauge that the de Donder gauge that they
adopted. The main reason for using this gauge is to relate
the classical instability to a negative mode. The negative mode can be
viewed as arising from a lower dimensional black hole solution, and is
therefore independent of the coordinates $z^i$. It therefore appears
simplest to adopt a gauge in which the perturbation is independent of
these coordinates.

Using the gauge \ref{eqn:pertgauge} and the above ansatz for the form of the
perturbation, the dilaton perturbation equation reduces to
\ba
\label{eqn:dileq}
 -\nabla^2 f &+& 2\beta \partial \bar{\phi} \cdot \partial f +
 H^{\mu\nu} \nabla_{\mu} \nabla_{\nu} \bar{\phi} \nonumber \\ &-& \beta H^{\mu\nu}
 \partial_\mu \bar{\phi} \partial_{\nu} \bar{\phi} 
 + \partial_{\mu} \bar{\phi} \nabla_{\nu} \left( H^{\mu\nu} -
 \frac{1}{2} H^{\rho}_{\rho} \bar{g}^{\mu\nu} \right) \nonumber \\
 &-& \frac{a}{2(n-1)!} e^{(\alpha+\beta) \bar{\phi}} \left( H^{\mu\nu}
 \bar{F}_{\mu\rho_1 \ldots \rho_{n-1}} \bar{F}_{\nu}{}^{\rho_1 \ldots
 \rho_{n-1}} - \frac{\alpha+\beta}{n} \bar{F}^2 f \right) = \lambda f,
\ea
where\footnote{
Note that $\mu^2 = \sum_i \mu_i \mu_i$, which should not be confused
with the parameter $\mu$ that appears in the black brane solution.}
$\lambda = -\mu^2$. The reason for writing the equations in this form
will soon become apparent. The $\mu\nu$ components of the metric 
perturbation equation
are
\ba
\label{eqn:metriceq}
 &-&\nabla^2 H_{\mu\nu} + 2 \nabla_{(\mu} \nabla^{\rho} H_{\nu) \rho} -
 \nabla_{\mu} \nabla_{\nu} H^{\rho}_{\rho} + 2 R_{\rho (\mu}
 H^{\rho}_{\nu)} + 2 R_{(\mu | \rho \sigma | \nu)} H^{\rho \sigma}
 \nonumber \\
 &-& \beta \left(2 \nabla_{(\mu} H^{\rho}_{\nu)} - \nabla^\rho
 H_{\mu\nu} \right) \partial_{\rho} \bar{\phi} + 2\beta \nabla_{\mu}
 \nabla_{\nu} f - 4(k+\beta^2) \partial_{(\mu} \bar{\phi}
 \partial_{\nu)} f \nonumber \\
 &+& \frac{1}{(n-1)!} e^{(\alpha + \beta) \bar{\phi}} \left( (n-1)
 H^{\rho \sigma} \bar{F}_{\mu\rho\lambda_1 \ldots \lambda_{n-2}}
 \bar{F}_{\nu\sigma}{}^{\lambda_1 \ldots \lambda_{n-2}} - \bar{F}_{\mu
 \lambda_1 \ldots \lambda_{n-1}} \bar{F}_{\nu}{}^{\lambda_1 \ldots
 \lambda_{n-1}} (\alpha+\beta) f \right) \nonumber \\
 &=& \lambda H_{\mu\nu}.
\ea
The $\mu i$ and $ij$ components of the metric perturbation equation
reduce to
\be
\label{eqn:Ydef}
 Y_\mu \equiv \nabla_{\nu} H^{\nu}_{\mu} - \beta H_{\mu}^{\nu}
 \partial_{\nu} \bar{\phi} - 2 f (k+\beta^2) \partial_{\mu} \bar{\phi}
 = 0,
\ee
\be
\label{eqn:Zdef}
 Z \equiv H^{\rho}_{\rho} - 2\beta f = 0.
\ee
It is straightforward to check that the trace of equation
\ref{eqn:metriceq} together with equations \ref{eqn:Ydef} and
\ref{eqn:Zdef} implies equation \ref{eqn:dileq}. Equation
\ref{eqn:Zdef} can be used to eliminate $f$ from equations
\ref{eqn:metriceq} and \ref{eqn:Ydef} and then the problem reduces to
solving equations \ref{eqn:metriceq} and \ref{eqn:Ydef}. Note that
equation \ref{eqn:Ydef} looks rather like a gauge condition. However,
this cannot be the case because a gauge has already been chosen, so
equation \ref{eqn:Ydef} should be regarded as an equation of motion.

GL looked for solutions with time dependence $e^{\Omega t}$. That
could be done here too, with equation \ref{eqn:Ydef}
used to express two of the four remaining unknown functions ($H^{tt}$, $H^{tr}$,
$H^{rr}$ and $K$) in terms of the other two.These could then be
substituted into equation \ref{eqn:metriceq} to give differential
equations for these two remaining unknowns. This appears to give
a set of equations in far fewer unknowns that those studied in
\cite{gregory:93, gregory:94}. This difference arises because the
gauge choice adopted above is not the same as the one used by GL.

This route will not be pursued here since the aim of this paper 
is to exhibit a connection between classical instability and Euclidean
negative modes. It was argued above that the necessary and sufficient
condition for a classical GL instability to exist is that there should
exist a static threshold unstable mode. For such a mode,
\be
 \Omega = H^{tr} = 0.
\ee

\subsection{Boundary conditions}

For a threshold unstable mode, the boundary conditions at the horizon of the
black brane are particularly simple. If the background metric is
written in the form \ref{eqn:metric} then the perturbed Lorentzian 
metric can be written
\be
\label{eqn:pertmetric}
 ds^2 = -U(r) (1 + \phi(r,z)) dt^2 + V(r)^{-1} (1+\psi(r,z)) dr^2 + 
 R(r)^2 (1 + k(r,z)) d\Omega^2 + dz^i dz^i,
\ee
where $\phi$, $\psi$ and $k$ are infinitesimal perturbations with
$z$-dependence $\exp(i \mu_i z^i)$. To examine regularity
of this perturbation on the future horizon, one can transform to
Eddington-Finkelstein \cite{hawking:73} coordinates. One finds that
the metric is regular at the horizon if, and only if, the perturbation
is bounded as $r \rightarrow r_+$ and
\be
\label{eqn:lorentzreg}
 \phi(r_+,z) = \psi(r_+,z).
\ee
At large $r$, the boundary condition is that the perturbation should
decay. 

\sect{Thermodynamic instability}

\label{sec:quantum}

In this section, it will be shown that a classical Lorentzian
threshold unstable mode exists if, and only if, the black hole
obtained by dimensional reduction of the black brane has a Euclidean
negative mode. 

\subsection{Wick rotation}

If a threshold unstable mode exists then it can be Wick rotated to
Euclidean signature (this will change the sign of $h_{tt}$). This gives
a solution to the equations describing perturbations around the Euclidean
signature black brane solution. As discussed in the introduction for the case of the
Schwarzschild black hole, the perturbation that will be related to the
Euclidean negative mode will not be the classical perturbation given
by $h_{MN}$ and $\delta \phi$ (which are proportional to 
$\exp(i \mu_i z^i)$), but rather the (off-shell) perturbation given by the
translationally invariant functions $H_{MN}$ and $f$. 
This perturbation satisfies
equations \ref{eqn:dileq} and \ref{eqn:metriceq}, which look rather
like eigenvalue equations for small quantum fluctuations around a
Euclidean black {\it hole} solution. Moreover, since $\lambda =
-\mu^2$ is negative, it appears that this particular fluctuation has a
negative eigenvalue and is therefore a Euclidean negative mode of the
black hole. In this section it will be shown that this is indeed
the case. This Euclidean perturbation will be referred to as a candidate
negative mode.

Since the background solution and the perturbation in question are both
translationally invariant, it is often convenient not to distinguish
between perturbations of the black brane and those of the black hole 
obtained by dimensional reduction. 
A negative mode of the black hole can be regarded as a
translationally invariant negative mode of the black brane. In order
to avoid having to work in two different conformal frames, the 
black hole will be studied in the frame in which its metric is given by simply
neglecting the flat directions of the black brane metric. Of
course, physical quantities such as masses and charges should be
calculated in the lower dimensional Einstein frame. However, if a
negative mode exists in one conformal frame then one must exist in
all other conformal frames since a change of variable in the Euclidean
path integral cannot change the number of negative modes.

It is necessary to show that the candidate negative mode is really a physical
negative mode. The first thing to check is that it satisfies the
appropriate boundary conditions for a quantum fluctuation around the
black hole. These are that the fluctuation should decay sufficiently
rapidly, and preserve regularity at $r=r_+$. The first of these
follows from the large $r$ behaviour of the Lorentzian solution, which
is clearly not changed by Wick rotation. The second can be seen by
Euclideanizing \ref{eqn:pertmetric} (i.e. setting $t = -i \tau$), 
dropping the $\exp(i \mu_i z^i)$ factors, and then defining a new  
radial coordinate $\rho$ near $r=r_+$ by
\be
 \rho = \frac{2 \sqrt{r-r_+}}{\sqrt{V'(r_+)}} \left( 1+ \frac{1}{2}
 \psi(r_+) \right).
\ee
In the new radial coordinate, the $\tau \rho$ part of the metric
becomes, near $r=r_+$,
\be
 d\rho^2 + \rho^2 \frac{U'(r_+) V'(r_+)}{4} \left( 1 - \psi(r_+) +
 \phi(r_+) \right) d\tau^2.
\ee
Regularity of the background solutions implies that $\tau \sim
\tau+\beta$ and equation \ref{eqn:reg} is satisfied. The perturbation
only preserves this regularity if it is bounded and satisfies
\be
 \psi(r_+) = \phi(r_+),
\ee
and this is clearly satisfied by the Wick rotated mode if it obeys the
Lorentzian boundary condition \ref{eqn:lorentzreg}.

\subsection{Negative modes}

The candidate negative mode can be written
\be
 \delta g_{\mu\nu} = H_{\mu\nu}, \qquad \delta g_{\mu i} = \delta
 g_{ij} = 0, \qquad \delta \phi = f.
\ee
$H_{\mu\nu}$ and $f$ satisfy the Wick rotated versions of 
equations \ref{eqn:dileq} and \ref{eqn:metriceq}, which can be written as
\be
\label{eqn:evaleq}
 {\bf G}(x) \int d^D x' {\bf O}(x,x') {\bf X}(x') = \sqrt{g} e^{-\beta
\bar{\phi}} \lambda {\bf X}(x)
\ee
where
\be
 {\bf X} = \left( \begin{array}{c} \delta \phi \\ \delta g_{MN}
\end{array} \right), \qquad {\bf O}(x,x') = \left( \begin{array}{cc}
\frac{\delta^2 I}{\delta \phi(x) \delta \phi(x')} & \frac{ \delta^2
I}{\delta \phi(x) \delta g_{PQ} (x')} \\ \frac{\delta^2 I}{\delta
g_{MN}(x) \delta \phi(x')} & \frac{\delta^2 I}{\delta g_{MN}(x) \delta
g_{PQ}(x')} \end{array} \right)  
\ee
and
\be
 {\bf G} = \left( \begin{array}{cc} 1 & \frac{2\beta}{D-2} g_{MN} \\
\frac{2\beta}{D-2} g_{PQ} & \, \left[2 g_{M(P} g_{Q)N} - \frac{2}{D-2}
\left(1-\frac{2 \beta^2}{D-2}\right) g_{MN} g_{PQ} \right]\end{array} \right).
\ee
The matrix ${\bf G} \equiv G_{IJ}$ can be regarded as the metric on the space of
fluctuations around a solution of the field equations, i.e., if the
fluctuation $(\delta g_{MN}, \delta \phi)$
is denoted schematically as $\psi_I$ then the kinetic
term for the fluctuation is $G^{IJ} \partial \psi_I \cdot \partial
\psi_J$. The presence of ${\bf G}$ in equation \ref{eqn:evaleq} is
required to balance up and down $I,J$ indices. 

Equation \ref{eqn:evaleq} is the equation governing quadratic
fluctuations around the Euclideanized black brane solution. 
So the candidate negative mode satisfies the correct
eigenvalue equation and the correct boundary conditions for quantum
fluctuations around the black brane, or black hole (by dimensional
reduction). The remaining thing to check is that it really is a {\it negative}
mode, i.e., a fluctuation that decreases the Euclidean
action. A mode with eigenvalue $\lambda$ has action
\be
\label{eqn:actioncont}
 \delta I = \lambda \int d^D x \sqrt{\bar{g}} e^{-\beta \bar{\phi}} {\bf X}^2,
\ee
where ${\bf X}^2 = {\bf X}^T {\bf G}^{-1}
{\bf X} = G^{IJ} X_I X_J$.  Expanding in terms of $f$ and $H_{\mu\nu}$
gives
\ba
 {\bf X}^2 &=& \left(1 - \frac{\beta^2 D}{D-2} \right) f^2 + \beta
 H^{\rho}_{\rho} f + \frac{1}{2} H^{\mu\nu} H_{\mu\nu} - \frac{1}{4}
 \left(H^\rho_\rho \right)^2 \nonumber \\
 &=& f^2 + \frac{1}{2} \tilde{H}^{MN} \tilde{H}_{MN} -
\frac{D-2}{4D} \left(\hat{H}^{M}_{M} \right)^2,
\ea
where $\hat{H}_{M N}$ is defined to be the $D$-dimensional {\it Einstein
frame} metric perturbation, with $\tilde{H}_{MN}$ its traceless part. 
It is clear that the metric ${\bf G}$ is not positive definite, but
has a negative eigenvalue associated with the trace of the Einstein
frame metric perturbation. In other words, the kinetic term for this
trace has the wrong sign. This is, of course, the conformal factor
problem \cite{gibbons:78}.

Equation \ref{eqn:actioncont} implies that modes for
which ${\bf X}^2 < 0$ everywhere are negative modes (i.e. decrease the action)
if $\lambda > 0$. These will be regarded as unphysical because they
arise from the conformal factor problem.
The modes with ${\bf X}^2 > 0$ everywhere are associated with directions
in the path integral along which the kinetic terms remain bounded, and
therefore if such a mode has $\lambda < 0$ then it represents a
physical non-conformal negative mode.

The candidate negative mode was obtained from a Lorentzian
solution with $Z=0$, so it has $H^{\rho}_{\rho} = 2 \beta f$. This implies
\be
 {\bf X}^2 = \left(1 - \frac{2 \beta^2}{D-2} + \frac{2 \beta^2}{n+2}
\right) f^2 + \frac{1}{2} H'^{\mu\nu} H'_{\mu\nu},
\ee
where $H'_{\mu\nu}$ is the traceless part of $H_{\mu\nu}$. A
sufficient condition for ${\bf X}^2 > 0$ is that the coefficient of
$f^2$ should be non-negative. Using
equations \ref{eqn:betadef}, \ref{eqn:framechoice} and
\ref{eqn:paramdef}, it is easy to show that this occurs if, and only if,
\be
\label{eqn:acond}
 a^2 \ge \frac{2(D-p-3)^2 (2-p)} {(D-2) (D-p)}.
\ee
This is automatically satisfied for $p \ge 2$. For $p=1$, the cases of
interest have to be considered individually. For a D$p$-brane, $D=10$
and $a = (p-3)/2$ and it follows that equation \ref{eqn:acond} is
satisfied for $p=1$. For the magnetic NS branes considered in
\cite{gregory:94}, one has $D=10$ and $a = (7-p)/2$, and equation
\ref{eqn:acond} is satisfied for $p=1$. Thus, for all the cases of
interest, the candidate negative mode is a non-conformal negative
mode.

\subsection{Proving the converse}

It has been shown that the classical Lorentzian threshold unstable mode
can be converted into a non-conformal Euclidean negative mode by dropping the $\exp(
i \mu_i z^i)$ factors and Wick rotating. If one accepts that the
existence of a non-conformal negative mode implies local thermodynamic
instability then it follows that the existence of a Gregory-Laflamme
instability implies that there is also a local thermodynamic instability.

To establish equivalence of classical and local thermodynamic
stability, it is necessary to prove the converse. As shown above, local
thermodynamic instability (negative specific heat) implies the
existence of a non-conformal negative mode for the lower dimensional
black hole solution, with some eigenvalue $\lambda < 0$.
This can be oxidized to give a negative mode of the black brane. 
This will satisfy the Euclidean versions of equations \ref{eqn:dileq} 
and \ref{eqn:metriceq} so it appears that this can now be Wick rotated and
multiplied by $\exp(i \mu_i z^i)$ (with $\mu^2 = -\lambda$) to obtain a
classical Lorentzian threshold unstable mode.

This conclusion is a little premature. 
Equations \ref{eqn:dileq} and \ref{eqn:metriceq} are indeed recovered
this way, but what about equations \ref{eqn:Ydef} and
\ref{eqn:Zdef}? At first sight, it is hard to see how these could
emerge from the negative mode since they are associated with the
classical equations of motion in the flat extra dimensions, 
which the negative mode does not ``know'' about. However,
if one takes the trace of equation \ref{eqn:metriceq} and combines it
with equation \ref{eqn:dileq}, one can derive
\be
\label{eqn:Zeq}
 -2 \nabla^2 Z + 2\beta \partial \bar{\phi} \cdot \partial Z + 2
 \nabla_\mu Y^{\mu} - 2\beta Y \cdot \partial \bar{\phi} = \lambda Z.
\ee
The combination $\nabla^\nu - \beta \partial^{\nu} \bar{\phi}$ times
equation \ref{eqn:metriceq} minus $2(k+\beta^2) \partial_{\mu}
\bar{\phi}$ times equation \ref{eqn:dileq} gives (after a long
calculation involving the background equations of motion, the Bianchi
identities and spherical symmetry)
\be
 \partial_{\mu} \left( -\nabla^2 Z + \beta \partial Z \cdot \partial
 \bar{\phi} + \nabla_{\mu} Y^{\mu} - \beta Y \cdot \partial \bar{\phi}
\right) = \lambda Y_{\mu}.
\ee
It follows that if $\lambda \ne 0$ then
\be
\label{eqn:YZ}
 Y_{\mu} - \frac{1}{2} \partial_{\mu} Z = 0,
\ee
i.e.,
\be
\label{eqn:gaugecond}
 \nabla_{\nu} \left( H^{\mu\nu} - \frac{1}{2} H^{\rho}_{\rho}
 \bar{g}^{\mu\nu} \right) = \beta H^{\mu\nu} \partial_\nu \bar{\phi} +
 2(k+\beta^2) f \partial_{\mu} \bar{\phi} + \beta \partial_{\mu} f.
\ee
This looks very much like a gauge condition on $H_{\mu\nu}$. One is,
of course, free to impose a gauge condition on $H_{\mu\nu}$ since this
is just the Euclidean metric perturbation for the black hole background, and 
no gauge condition has yet been imposed on this perturbation. 
However, it seems rather odd that a gauge condition should emerge from 
the equations of motion! The resolution of this puzzle is simply that, for $\lambda
 \ne 0$, the above equations are not equations of
motion, and in fact the right hand side of the
eigenvalue equation is {\it not} gauge invariant. A similar phenomenon
would occur in Maxwell theory: the equation
$\partial_{\nu} \delta F^{\mu\nu} = \lambda \delta A^\mu$ implies $\partial
\cdot \delta A = 0$ when $\lambda \ne 0$. Therefore, this eigenvalue
equation is only valid Lorentz gauge. In a different gauge, the right hand
side of the eigenvalue equation would be different.

Further progress can be made by using equation \ref{eqn:YZ} to
eliminate $Y_\mu$ from equation \ref{eqn:Zeq}, giving
\be
 - \nabla^2 Z + \beta \partial Z \cdot \partial \bar{\phi} = \lambda Z,
\ee
which implies
\be
 \int d^D x \sqrt{\bar{g}} e^{-\beta \bar{\phi}} (\partial Z)^2 =
\lambda \int d^D x \sqrt{\bar{g}} e^{-\beta \bar{\phi}} Z^2.
\ee
The left hand side of this equation is positive. A normalizable
negative mode would give a negative right hand side, so it follows that
such a mode must have $Z=0$. Using equation \ref{eqn:YZ} it follows
that a normalizable negative mode must have $Z = Y_{\mu} = 0$, so equations
\ref{eqn:Ydef} and \ref{eqn:Zdef} are indeed satisfied. This completes
the proof that that the threshold classical Lorentzian 
unstable mode in the gauge \ref{eqn:pertgauge}
is obtained from the negative mode in the gauge \ref{eqn:gaugecond} 
simply by Wick rotation and multiplying by $\exp(i \mu_i z^i)$ 
with $\mu^2 = -\lambda$. 

\sect{Conclusions}

It has been explained why there should be a relationship between
classical and thermodynamic stability of black branes, as conjectured
by Gubser and Mitra. Various assumptions were made which means that
the above discussion falls well short of a rigorous proof. It is
convenient to summarize the arguments used above, to emphasize where
these assumptions were made. 

A) Local thermodynamic stability of a magnetically charged black
brane is equivalent to positivity of the
specific heat of the black hole obtained by dimensional reduction. If
the black brane is locally thermodynamically unstable then so is the
black hole, so it will have a negative mode. It was assumed 
that this is an s-wave, but this is probably acceptable 
since it seems very likely that higher partial waves would have higher eigenvalues.
This negative mode can be oxidized to give a negative mode of the black brane.
If a suitable gauge is chosen then this negative mode can be converted into a
classical threshold unstable mode as described above. Such a mode
would be expected to separate stable short wavelength fluctuations of
the black brane from unstable long wavelength fluctuations, so the
existence of such a mode indicates that a classical Gregory-Laflamme instability
exists. So local thermodynamic instability implies classical
instability.

B) For the converse, if a black brane is classically unstable then
the instability is expected to occur only at long wavelengths. Therefore
there should exist a threshold unstable mode that is independent of
time. Provided this mode is of the longitudinal s-wave form discussed by Gregory and
Laflamme, then it is possible to choose a gauge in which the
components of the metric perturbation along the spatial worldvolume
directions vanish. Dropping the $\exp(i \mu_i z^i)$ factors and Wick
rotating then gives an s-wave negative mode of the black hole. This
implies the existence of a local thermodynamic instability of the
black hole, assuming that the results of \cite{prestidge:00b} 
extend to the black holes considered here. It then follows that the 
black brane must also be locally thermodynamically unstable. 

The assumptions involved in proving A) that local thermodynamic
instability implies classical instability seem rather mild. The
assumptions involved in proving the converse B) are less obvious, so it
would be interesting to see if these could be given better
justification. To see why proving B) rigorously is hard, consider the
case when the black brane has positive specific heat. Proving B) in
this case amounts to proving that the black brane is classically
stable, which involves considering {\it all} fluctuations around the
black brane, not just those of the GL form. 

If a black brane is sufficiently far from extremality then it should
behave much like an uncharged black brane and therefore have negative
specific heat. It then follows from argument A) above that the black brane
will be classically unstable. As the brane approaches extremality,
argument A) establishes that the brane will be classically unstable as
long as the specific heat remains negative. So argument A) extends the
results of GL to any magnetically charged black brane with negative
specific heat. If argument B) is correct then the instability should 
vanish when the specific heat becomes positive.

It has been explained that the reason GL found
an instability that persisted all the way down to extremality is that
the branes they studied always have negative specific heat. Had
they chosen to work with branes whose specific heat became positive
near extremality then, according to B), they would not have
found an instability near extremality.

For a black brane of a type II supergravity theory, with the exception of
the D3-brane, it is always possible to dualize the field strength so
that the brane is magnetically charged and therefore falls within the
class of solutions considered in this paper. The non-extremal NS5, D5
and D6 branes always have negative specific heat so A) implies that
they are always classically unstable. 
However, the F1, D1, D2 and D4-branes have a specific
heat that becomes positive near extremality. A) establishes that they
should be classically unstable when their specific heat is negative,
and B) implies that they should become classically stable near
extremality.  Note that the F-string
and D-string have the same stability properties, as do the NS5 and
D5-branes, 
as one would expect from the $SL(2,R)$ symmetry of IIB supergravity.
The major omission from this paper is, of course, the black D3-brane. I
think that it is likely that the above analysis could be extended to cover
this case, as well as the other dyonic black brane solutions given in
\cite{duff:96}. If this is the case then the black D3-brane should be
stable near extremality.

The M2 and M5-branes are not covered by the analysis above but one can
infer their stability properties from the behaviour of the F-string
and D4-branes, which are obtained by dimensional reduction on a
worldvolume direction. If the higher dimensional brane were
classically unstable to a perturbation of the form considered above
then one could reduce in a direction orthogonal to $\mu_i$ and the
instability would persist in the lower dimensional solution. Since a
GL instability is not expected for the near-extremal F-string or
D4-brane, this implies that the near extremal M2 and M5-branes must
be classically stable, in agreement with the specific heat analysis.

The GM conjecture is supposed to cover any black brane with a
non-compact translational symmetry. The branes considered in this
paper were all assumed to have flat worldvolumes. However, it might be
possible to extend the analysis to branes with Ricci flat 
spatial worldvolume metrics of the ``Kaluza-Klein'' form
\be
 ds^2 = e^{\alpha \phi} (d\tau + A )^2 + e^{\beta\phi} \gamma_{ij} dy^i dy^j,
\ee
where $A$ is a $1$-form independent of $\tau$, the coordinate along the 
non-compact direction. The classical instability
would presumably involve fluctuations proportional to $\exp (i \mu
\tau)$. 

Perhaps the most striking result of this paper is that it is possible
to show that many black branes are classically unstable without
resorting to the arduous numerical analysis of \cite{gregory:93,
gregory:94}. Instead one just has to calculate the specific heat. So a
simple thermodynamic calculation replaces a difficult classical
calculation. This can be contrasted with black hole thermodynamics, 
where the temperature and entropy are, in principle, obtained by
a long calculation \cite{hawking:75}, but in practice are determined
by simple classical quantities, namely the surface gravity and horizon
area. Understanding these connections between classical and quantum
properties of black holes remains one of the most interesting problems
in quantum gravity.

\bigskip

\begin{center}{\bf Acknowledgements}\end{center}

I would like to thank Fay Dowker, Steven Gratton, Ruth Gregory, Stephen
Hawking, Tim Prestidge and Simon Ross for illuminating discussions, 
Fay Dowker and Ruth Gregory for comments on a draft of this paper, 
Roberto Emparan for comments on the first preprint version,
and Tim Prestidge for allowing me to use unpublished results. This
work was funded by PPARC.

\end{document}